\documentclass{PoS}

\usepackage{graphicx}  
\usepackage{amsmath,amssymb}  
\usepackage{caption}
\usepackage{subcaption}
\usepackage{gensymb}
\usepackage{multicol}

\title{VHE Analyses of Long-Term Low-Flux-State Observations by VERITAS of Intermediate-Frequency-Peaked BL Lacertae Sources: 3C 66A and W Comae}

\ShortTitle{VHE Analyses of Long-Term Low-Flux-State Observations by VERITAS}

\author{\speaker{Juliana Vievering} for The VERITAS Collaboration\thanks{veritas.sao.arizona.edu} \\
        University of Minnesota\\
        E-mail: \email{vieverin@physics.umn.edu}}

\abstract{Intermediate-frequency-peaked BL Lacertae objects (IBLs) are a class of blazars characterized by a spectral energy distribution (SED) with a lower-energy synchrotron peak than the majority of extragalactic sources detected by ground-based imaging atmospheric Cherenkov telescopes (IACTs). Consequently, the peak gamma-ray flux falls outside the very-high-energy regime (VHE, >100~GeV) covered by IACTs such as VERITAS, making IBLs difficult to detect except during infrequent episodes of elevated flux. However, the study of these sources in a low-flux state is essential for developing a complete understanding of the blazar paradigm. We present the results of VHE analyses of long-term low-flux-state observations completed for two IBL sources: 3C 66A and W Comae. For both sources, data from VERITAS were analyzed for the VHE regime. The study of 3C 66A extends from 2007 to 2015, resulting in a 12 standard deviation ($\sigma$) detection from $\sim$61 observing hours. Analysis of W Comae from 2010 to 2014, totaling $\sim$39 hours, resulted in a 6$\sigma$ low-flux-state detection. We report on the results from these VHE analyses and describe contemporaneous multiwavelength data to be used in further analyses. We comment on how these low-flux-state IBL detections fit within the context of the blazar paradigm.}

\FullConference{The 34th International Cosmic Ray Conference,\\
		30 July- 6 August, 2015\\
		The Hague, The Netherlands}

\begin{document}
\section{Introduction}
Blazars are a type of active galactic nuclei (AGN) with collimated relativistic jets, specifically ones where the jets are aligned with our line of sight. Blazars are some of the most energetic astrophysical objects in the Universe and are found observationally to have a characteristic two-peak spectral energy distribution (SED). The lower-frequency peak in photon flux is associated with synchrotron radiation resulting from relativistic electrons accelerating in strong magnetic fields. In leptonic models, the second peak is thought to be associated with the inverse-Compton (IC) effect, with the seed photons being provided by synchrotron radiation (synchrotron self-Compton, SSC), external photon fields (external Compton, EC), or some combination of the two e.g. \cite{blazar_spectrum}. \\
\indent Historically, blazars have been classified as one of two types based on spectral features: flat-spectrum radio quasars (FSRQs) or BL Lacertae objects (BL Lacs). FSRQs exhibit strong optical emission lines while BL Lacs generally appear to be featureless. FSRQs are generally thought to host more efficient accretion disks and stronger jets resulting in lower synchrotron peak frequencies ($\nu_{s}^{peak}$), while BL Lacs typically show weaker jets driven by inefficient accretion and cover a range of $\nu_{s}^{peak}$ categorized accordingly: high-, intermediate-, or low-frequency-peaked BL Lacs (HBLs, IBLs, or LBLs) \cite{bl_lacs}. However, it has been argued that some of these distinctions are due to sample selection effects and the wide variety of blazar observables could be accommodated through association of BL Lacs and FSRQs with low-excitation (LERGs)/Fanaroff-Riley (FR) I and high-excitation (HERGs)/FR II radio galaxies \cite{Giommi2012, Buttiglione2010}. \\
\indent Thus, one of the outstanding problems within blazar observations is the seeming differentiation of blazars into sub-classes as a function of observables such as position of $\nu_{s}^{peak}$, the overall source luminosity and the dominance of the IC peak luminosity relative to the synchrotron peak ("Compton dominance"). These observables can be related to intrinsic physical properties of the system such as the maximum particle acceleration energy, the jet power or accretion mode and the relative fraction of the external photon fields to the seed photon density.  The synchrotron peak is observed to fall in a range of frequencies covering nearly five orders of magnitude, from infrared to x-ray wavebands and empirically, the Compton dominance is anti-correlated with $\nu_{s}^{peak}$ \cite{bl_lacs}. In the leptonic model paradigm, the FSRQ, LBL, IBL, HBL sequence requires a decreasing external photon field component to model the decreasing Compton dominance; FSRQ jets are in the most gas enriched environments requiring the highest EC component while HBLs are typically modeled satisfactorily with a simple SSC model requiring no EC component \cite{BoettcherDermer2002}. IBLs are transition objects and so far have been detected only in flaring states in which SED models have required an EC component \cite{wcom_flare}. If indeed, Compton dominance can be related to underlying physical observables of the blazar system, then it is important to understand how the system changes between a quiescent and flaring state. \\
\indent Thus, to build a more robust understanding of the nature of blazars, it is essential to study these transition sources in greater depth. VERITAS has detected many HBLs, but IBL and LBL sources are particularly challenging for VHE observatories to detect outside infrequent periods of elevated flux. Since the two-peak SED shifts to lower frequencies for IBL/LBLs, the VHE regime falls beyond the inverse-Compton peak, leading to very limited observed flux. Therefore, these sources require long-term observations to be detected in a low-flux state. Although detections in high-flux states (flares) for these sources already exist, it cannot be assumed that IBLs will exhibit the same spectral behavior in a low-flux state; there is already ample evidence of spectral variability for HBLs between high- and low-flux states \cite{HBLspecvar,421specvar}. In particular, it is important to ascertain whether the "quiescent" or low-flux-state SED for IBLs requires an external photon field to appropriately model the IC peak. This information can be combined with other observables to yield a more comprehensive understanding of blazars overall. \\
\indent These proceedings describe the results of the VHE analyses of long-term low-flux-state observations for two IBL sources: 3C 66A and W Comae. Section 2 provides a description of the VERITAS observatory and summarizes the past analyses of these two sources in both the high- and low-flux states. Section 3 describes highlights of the low-flux-state analysis of 3C 66A, and Section 4 details the analysis of W Comae, presenting a low-flux-state spectrum for this source for the first time ever. Finally, in Section 5 we offer our conclusions from the analyses thus far and describe future work to be done with these sources.
\begin{figure}[b]
\centering
\begin{subfigure}[b]{.4\textwidth}
  \centering
  \includegraphics[width=\textwidth]{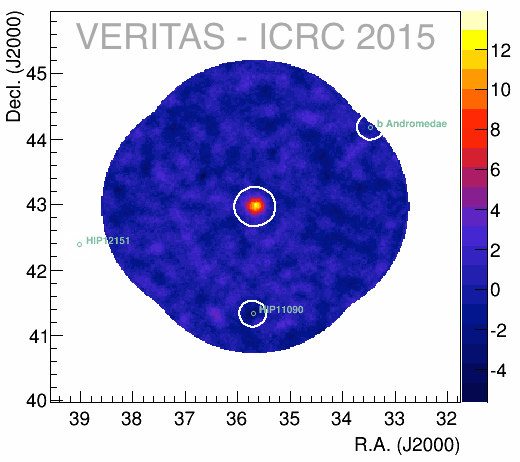}
  \caption{3C 66A}
  \label{fig:sm2}
\end{subfigure}  
\begin{subfigure}[b]{.4\textwidth}
  \centering
  \includegraphics[width=\textwidth]{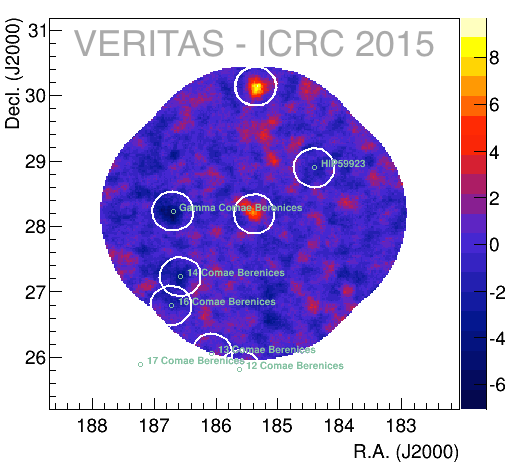}
  \caption{W Comae}
  \label{fig:sm1}
\end{subfigure}
\caption{\footnotesize \emph{Significance maps (standard cuts) plotted in RA and Dec (degrees) with significance indicated by color bar. The white circles indicate regions excluded from the background estimation because they contain stars or gamma-ray sources. (a) The analysis for 3C 66A excluded stars brighter than magnitude 7. (b) For W Comae, stars with magnitude brighter than 6 were excluded as well as the gamma-ray source 1ES 1218+304.}}
\label{fig:sm}  
\end{figure}
\section{The VHE Detections of 3C 66A and W Comae}
VERITAS is a ground-based array of four imaging atmospheric Cherenkov telescopes (IACTs) at the Fred Lawrence Whipple Observatory in southern Arizona. Each telescope is comprised of a segmented 12m primary mirror and a camera made of 499 photomultiplier tubes (PMTs). The VERITAS telescopes have a $3.5\degree$ field of view and are optimized for the energy range 85 GeV - 30 TeV \cite{weekes}. VERITAS was constructed in phases; the initial configuration of the four-telescope array (2007-2009), called the "old array" (OA), was changed to the "new array" (NA) in the fall of 2009 when one of the telescopes was relocated to a more optimal position for improved sensitivity. Prior to the fall of 2012, the PMTs were replaced, leading to the "upgraded array" (UA), which is in current use. The sources discussed in these proceedings were observed at a 0.5$\degree$ offset from the camera center for purposes of simultaneous background estimation \cite{analysis}. For more details on VERITAS, see \cite{veritas_details}. \\
\indent A VHE detection of 3C 66A ($0.3347\textless z \textless 0.41$) \cite{3c66a_redshift} was first reported by the Crimean Astrophysical Observatory in 1998 \cite{3c66a_CAO}, and a detection of the source by VERITAS was first reported in 2008. During an elevated flux state observed for several days at the beginning of October 2008, the source reached over 8$\sigma$ significance, corresponding to an integral flux ($\textgreater$100$~$GeV) that was $\sim$10\% the flux of the Crab Nebula \cite{3c66a_atel}. VERITAS first reported a low-flux-state detection (5.4$\sigma$) of 3C 66A in $\sim$17 hours of observations from January 2010 to June 2012 \cite{tommy}. \\
\indent A detection of W Comae ($z=0.102$) was first reported in the VHE band by VERITAS after an elevated flux state in March 2008. During the two nights of highest activity, W Comae was observed at an integral flux ($\textgreater$200$~$GeV) equivalent to 9\% flux from the Crab Nebula \cite{wcom_detection}. Shortly after, in June 2008, W Comae was observed in an even higher flux state (see table \ref{stats}), with integral flux ($\textgreater$200$~$GeV) corresponding to 25\% Crab \cite{wcom_flare}. VERITAS was also first to report a low-flux-state detection of W Comae (5.6$\sigma$) through analysis of $\sim$31 hours of data January 2010 to June 2012 \cite{tommy}. In these proceedings, we extend the low-flux-state analyses for both 3C 66A and W Comae. 
\begin{figure}[t]
\centering
\begin{subfigure}{0.49\textwidth}
  \centering
  \includegraphics[width=\textwidth]{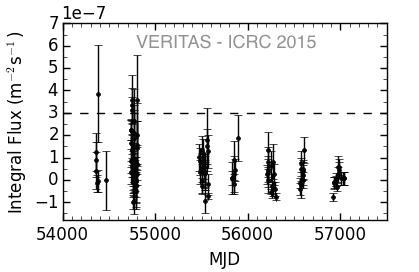}
  \caption{3C 66A}
  \label{fig:lc2}
\end{subfigure}  
\begin{subfigure}{0.49\textwidth}
  \centering
  \includegraphics[width=\textwidth]{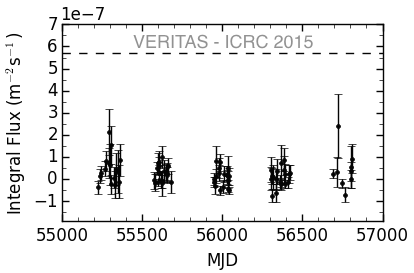}
  \caption{W Comae}
  \label{fig:lc1}
\end{subfigure}
\caption{\footnotesize \emph{Light curves of low-flux-state analysis for each source, binned by day and plotted against MJD. Observing seasons for VERITAS run from September to June. The integral flux ($m^{-2}s^{-1}$) was calculated above 200 GeV.  (a) Data for 3C 66A ranges from September 2007 to January 2015. The dotted line represents the average integral flux ($\textgreater$200 GeV) from the high-flux state (October 2008 flare) of 3C 66A. (b) Data for W Comae ranges from January 2010 to May 2014, dotted line represents the average integral flux ($\textgreater$200 GeV) from the June 2008 flare \cite{wcom_flare}.}}
\label{fig:lc}  
\end{figure}
\section{Highlights of 3C 66A Analysis}
The most recent analysis of 3C 66A, using VERITAS observations from September 2007 to January 2015, consists of $\sim$61 hours (livetime) in a low-flux state and $\sim$6 hours (livetime) in a high-flux state, after data quality cuts to remove observations with poor weather or hardware issues. The high-flux state (October 2008 flare) was distinguished from the low-flux state by showing that the probability of randomly seeing an integral flux level significantly higher than the mean flux three days in a row is negligible. Analysis of this source utilized a combination of soft ($\textless$200$~$GeV) and standard ($\textgreater$200$~$GeV) cuts, determined by studying data from the Crab Nebula scaled to 10\% flux as a toy model. Soft cuts allow for events of lower minimum size (as measured by digital counts) than the standard cuts and also use a larger angular region around the source to account for poorer angular resolution at this lowered threshold. \\
%
\indent A position-based analysis of VERITAS observations has shown that the center of the excess is associated with 3C 66A, ruling out radio galaxy 3C 66B at 95\% confidence. Analysis of the low-flux state resulted in a 12$\sigma$ detection. The photon spectrum for the low-flux state was found to be best fit with a log-parabola function of the form $dN/dE=C(E/E_0)^{-(\Gamma+\beta log(E/E_0))}$ where C is the normalization constant, $E_0$ is the decorrelation energy, $\Gamma$ is the photon index, and $\beta$ is the curvature parameter. The decorrelation energy, $E_0=0.27~$TeV in this analysis, represents the energy at which the calculated flux is least affected by the choice of spectral index. See Table \ref{3C66A} for the other parameters and Figure \ref{fig:sp2} for the low-flux-state spectrum. Since a log-parabola model is preferred over a power law at 4.1$\sigma$, we conclude that there is significant curvature to the photon spectrum. From the results of the fits, we also note that the photon index $\Gamma$ and curvature parameter $\beta$ for the low- and high-flux states agree within statistical uncertainty, showing that there is currently no observable change of shape in the VHE spectrum between states. A more detailed description of the analysis of 3C 66A and conclusions drawn will appear in an upcoming publication from the VERITAS collaboration. 
\begin{table}[t]
\begin{center}
\begin{tabular}{lcccccc}
   \hline
   \textbf{\small 3C 66A} & \footnotesize Expos. & \footnotesize Signif. & \footnotesize Normalization $C$ & \footnotesize Photon Index & \footnotesize Curvature & \footnotesize Integral flux \\
    & \footnotesize (hrs) & \footnotesize $\sigma$ & \footnotesize (m$^{-2}$s$^{-1}$TeV$^{-1}$) & \footnotesize $\Gamma$ & \footnotesize $\beta$ & \footnotesize (m$^{-2}$s$^{-1}$) \\
  \hline
  \small (a) Low & \small 61 & \small 12 & \small $(3.62\pm0.33)\times10^{-7}$ & \small $4.67\pm0.24$ & \small $3.65\pm1.24$ & \small $(7.08\pm0.74)\times10^{-8}$ \\
  \small (b) High & \small 6 & \small 13 & \small $(1.40\pm0.16)\times10^{-6}$ & \small $5.18\pm0.34$ & \small $2.29\pm1.72$ & \small $(2.98\pm0.37)\times10^{-7}$ \\
  \hline 
\end{tabular}
\caption{\footnotesize \emph{Statistics of low- and high-flux states of 3C 66A. Exposure is the total livetime in hours after data quality cuts. Analysis for both states uses a combination of soft cuts ($\textless$200 GeV) and standard cuts ($\textgreater$200 GeV). A log-parabola function was used to model the data. The lower limit for the integral flux is 200 GeV; the integral flux corresponds to $\sim$3$\%~$Crab for the low-flux state and $\sim$12$\%~$Crab for the high-flux state. All errors shown are statistical.}}
\label{3C66A}
\end{center}
\end{table}
%
%
%
\section{Detailed VHE Analysis of W Comae in Low-Flux State}
\begin{figure}[t]
\centering
\begin{subfigure}{.49\textwidth}
  \centering
  \includegraphics[width=\textwidth]{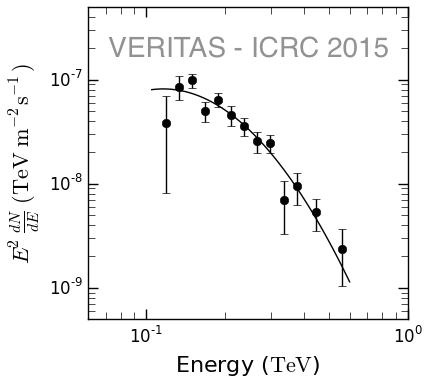}
  \caption{3C 66A}
  \label{fig:sp2}
\end{subfigure} 
\begin{subfigure}{.49\textwidth}
  \centering
  \includegraphics[width=\textwidth]{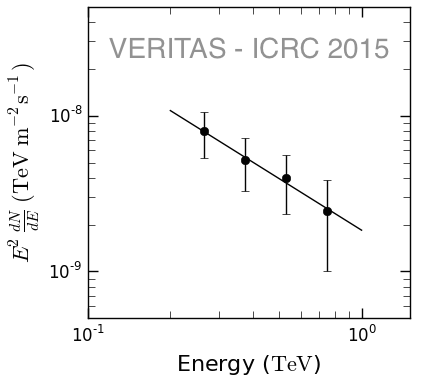}
  \caption{W Comae}
  \label{fig:sp1}
\end{subfigure} 
\caption{\footnotesize \emph{Low-flux-state VHE spectra of W Comae and 3C 66A, shown in the $E^2 dN/dE$ representation (TeV m$^{-2}$s$^{-1}$). Bins with significance $\textgreater$2$\sigma$ and at least 5 signal events after background subtraction are plotted. All errors plotted are statistical. (a) 3C 66A is detected in a low-flux state at 12$\sigma$ using soft ($\textless$200$~$GeV) and standard ($\textgreater$200$~$GeV) cuts with 61 hours of observations. A log-parabola function is used to model the data. (b) W Comae is detected in low-flux state at 6.2$\sigma$ using standard cuts with 39 hours of livetime. A power law function is used to model the data.}}
\label{fig:sp}  
\end{figure}

Beyond the previous low-flux-state detection of W Comae in 2012, this analysis includes two seasons of observations (2012/13 and 2013/14) from the upgraded VERITAS array, resulting in $\sim$39 hours of total livetime after data quality cuts. All observations by VERITAS of W Comae over the span of 2010-2014 are defined as the low-flux state, which was determined by comparing the integral flux of the high-flux state (June 2008 flare) to the daily light curve from 2010 to 2014 (see Figure \ref{fig:lc2}). \\
\indent Two independent analyses of W Comae were performed using standard cuts and produced consistent results. The VHE significance map (Figure \ref{fig:sm1}) shows the VERITAS field a view in the vicinity of W Comae. A reflected region background model was used, where background is estimated from circular regions of equal size and at equal distance from the camera center as the source (0.5$\degree$ offset) \cite{analysis}.  From this analysis, observations yielded 164 excess events ($N_{on}=883$ and $N_{off}=6933$, average normalization factor $\alpha=0.1014$), resulting in a 6.2$\sigma$ detection and integral flux ($\textgreater$200$~$GeV) of $(2.81\pm0.53)\times10^{-8}$ m$^{-2}$s$^{-1}$ from $\sim$39 hours of livetime. The first ever low-flux-state VHE spectrum of W Comae is presented in Figure \ref{fig:sp1}. \\ 
\begin{table}[b]
\begin{center}
\begin{tabular}{lccccc}
  \hline
  \textbf{\small W Comae} & \footnotesize Exposure & \footnotesize Significance & \footnotesize Normalization $C$ & \footnotesize Photon Index & \footnotesize Integral Flux \\
   & \footnotesize (hrs) & \footnotesize $\sigma$ & \footnotesize (m$^{-2}$s$^{-1}$TeV$^{-1}$) & \footnotesize $\Gamma$ & \footnotesize (m$^{-2}$s$^{-1}$) \\
  \hline
  (a) \small Low & \small 39 & \small 6 & \small $(6.29\pm1.41)\times10^{-8}$ & \small $3.1\pm0.6$ & \small $(2.81\pm0.53)\times10^{-8}$ \\
  (b) \small High & \small 4 & \small 10 & \small $(6.5\pm0.9)\times10^{-7}$ & \small $3.68\pm0.22$ & \small $(5.7\pm0.6)\times10^{-7}$ \\
  \hline
\end{tabular}
\caption{\footnotesize \emph{Statistics of low-flux and high-flux (June 2008 flare) states of W Comae. $E_0=0.4~$TeV for June 2008 flare \cite{wcom_flare}. The data is fit with a power law model for both states. The lower limit for integral flux is 200 GeV; the integral flux corresponds to $\sim$1$\%~$Crab for the low-flux state and $\sim$25$\%~$Crab for the high-flux state. All errors shown are statistical. }}
\label{stats}
\end{center}
\end{table}
\indent The spectrum was fit with a power law model of the form $dN/dE=C(E/E_0)^{-\Gamma}$, where $C$ is the normalization constant, $E_0$ is the decorrelation energy ($E_0=0.32~$TeV in this case), and $\Gamma$ is the photon index. The best fit of the data using this model results in $C=(6.29\pm1.41_{stat})\times10^{-8}$ m$^{-2}$s$^{-1}$TeV$^{-1}$ and $\Gamma=3.1\pm0.6_{stat}$. We note that the value of the photon index in the low-flux state for W Comae is in agreement with the photon index for the June 2008 flare ($\Gamma=3.68\pm0.22_{stat}$) \cite{wcom_flare} within statistical uncertainty, showing that we currently find no significant change in the spectral shape between states in the VHE waveband. \\
\indent Future work with W Comae will continue to provide greater insight into the underlying physics behind IBL sources. This includes extending the analysis both within the VHE waveband and to other wavelengths. In the VHE waveband, we intend to integrate data from the old array (2007/08 and 2008/09 seasons) and also the most recent season of data from the upgraded array (2014/15). Analysis of the most recent season is already underway; our preliminary results show W Comae to be unusually faint in VHE during 2014/15. \\
%
%
\indent In addition to continued work with VERITAS data, we are in the process of gathering publicly available multiwavelength data from radio to HE gamma rays for the purpose of constructing the full multi-band SED most representative of the low-flux state. In the HE regime, we have already begun an analysis of W Comae using data from the \emph{Fermi} Large Area Telescope (LAT). 
As a \emph{Fermi} source of interest, W Comae is one of many blazars also monitored by \emph{Swift} in the x-ray through optical wavebands (\emph{Swift} XRT and UVOT). Upon initial study, \emph{Swift} public x-ray data shows a decline in integral flux over the course of the most recent season of VERITAS data (January-April 2015), matching our preliminary VHE results. Several optical observatories have long-term data on W Comae, including the Catalina Sky Survey, the FLWO 48'' Telescope, and Tuorla, and a combination of these datasets will provide contemporaneous observations in the optical regime. In radio, W Comae is observed by the Owens Valley Radio Observatory (OVRO) at $15~$GHz approximately twice each week, and thus we expect there to be contemporaneous data with the VHE observations. 
We will continue to investigate this dataset and query other radio observatories for more data to include in a low-flux-state SED. Study of this full SED will provide insight into the spectral variability of W Comae between high- and low-flux states and also can lead to understanding of the intrinsic jet strength. 
%
%
%
\section{Conclusions}
Through long-term, low-flux-state analyses of 3C 66A and W Comae, we contribute the VHE portion of a future SED for each of these IBL sources in their low-flux states. VERITAS observations of 3C 66A totaling $\sim$61 hours of live time in the low-flux state led to a 12$\sigma$ detection. A log-parabola model was found to better fit the spectrum than a power law model, showing significant curvature in the VHE spectrum. Observations of W Comae totaling $\sim$39 hours led to a 6.2$\sigma$ detection, allowing for a VHE spectrum in the low-flux state to be derived for the first time. We see no significant curvature in the VHE spectrum for W Comae, with the data being best fit to a power law model. For both sources, there is no significant difference in the spectral shape between the low- and high-flux states in VHE. Continued observations of these sources in a low-flux state will allow for greater resolution in the VHE spectra and better constrained parameters. \\
\indent These results provide the initial step towards multiwavelength analyses. The VHE spectrum in combination with other wavelengths will define the shape and locations of both the synchrotron and inverse-Compton peaks, providing insight on the nature of IBLs in multiple ways. It has been shown that there is a shift in the SED of HBL sources during flaring states, and so we can determine whether this effect also exists for IBLs once a low-flux-state SED is produced. Furthermore, the relative luminosities of the synchrotron and inverse-Compton peaks can offer information on the strength of external photon fields, which could provide evidence for whether or not IBLs are driven by weak or strong jets. The study of IBLs as transition objects in synchrotron peak frequency provides an essential step towards developing greater understanding of the overall blazar paradigm. \\

\noindent \textbf{Acknowledgements} \\
{\footnotesize This research is supported by grants from the U.S. Department of Energy Office of Science, the U.S. National Science Foundation and the Smithsonian Institution, and by NSERC in Canada. We acknowledge the excellent work of the technical support staff at the Fred Lawrence Whipple Observatory and at the collaborating institutions in the construction and operation of the instrument. The VERITAS Collaboration is grateful to Trevor Weekes for his seminal contributions and leadership in the field of VHE gamma-ray astrophysics, which made this study possible.}

\end{document}